\documentclass[12pt,preprint]{aastex}

\begin{document}

\newcommand{\zdot}{\makebox[0pt][l]{.}}
\newcommand{\up}[1]{\ifmmode^{\rm #1}\else$^{\rm #1}$\fi}
\newcommand{\dn}[1]{\ifmmode_{\rm #1}\else$_{\rm #1}$\fi}
\newcommand{\upd}{\up{d}}
\newcommand{\uph}{\up{h}}
\newcommand{\upm}{\up{m}}
\newcommand{\ups}{\up{s}}
\newcommand{\arcd}{\ifmmode^{\circ}\else$^{\circ}$\fi}
\newcommand{\arcm}{\ifmmode{'}\else$'$\fi}
\newcommand{\arcs}{\ifmmode{''}\else$''$\fi}

\pagestyle{plain}

\def\thefootnote{\fnsymbol{footnote}}

\title{Difference Image Analysis of the OGLE-II Bulge Data. III.\\
Catalog of 200,000 Candidate Variable Stars. \footnotemark[1]}

\footnotetext[1]{Based on observations obtained with the 1.3-m
Warsaw telescope at Las Campanas Observatory of the Carnegie Institution
of Washington}

\author{P. R. Wo\'zniak$^{1,2}$, A. Udalski$^3$, M. Szyma\'nski$^3$,
M. Kubiak$^3$, G. Pietrzy\'nski$^{3,4}$, I. Soszy\'nski$^3$, K. \.Zebru\'n$^3$
}

\affil{$^1$ Princeton University Observatory, Princeton, NJ 08544, USA}
\affil{$^2$ Los Alamos National Laboratory, MS-D436, Los Alamos, NM 87545, USA}
\affil{e-mail: wozniak@lanl.gov}

\affil{$^3$ Warsaw University Observatory, Al. Ujazdowskie 4, 00-478 Warszawa,
Poland} 
\affil{e-mail: (udalski,msz,mk,pietrzyn,soszynsk,zebrun)@astrouw.edu.pl}

\affil{$^4$ Universidad de Concepci\'on, Departamento de Fisica,
Casilla, 160-C, Concepci\'on, Chile}
\affil{e-mail: pietrzyn@hubble.cfm.udec.cl}

\begin{abstract}

We present the first edition of a catalog of variable stars from
OGLE-II Galactic Bulge data covering 3 years: 1997--1999. Typically
200--300 $I$ band data points are available in 49 fields between $-11$
and $11$ degrees in galactic longitude, totaling roughly $11$ square
degrees in sky coverage. Photometry was obtained using the Difference
Image Analysis (DIA) software and tied to the OGLE data base with
the DoPhot package. The present version of the catalog comprises
221,801 light curves. In this preliminary work the level of
contamination by spurious detections
is still about 10\%. Parts of the catalog have only crude calibration,
insufficient for distance determinations. The next, fully calibrated,
edition will include the data collected in year 2000.
The data is accessible via FTP. Due to the data volume, we also
distribute DAT tapes upon request.

\end{abstract}

\section{Introduction}

\label{sec:intro}

The main goal of the Optical Gravitational Lensing Experiment
(OGLE, Udalski, Kubiak \& Szyma\'nski 1997) is to search for
microlensing events. Observationally, these events are basically
a rare type of an optical variable, therefore it came as no surprise
that after several years microlensing experiments have an exceptional
record of variability in terms of the number of objects and epochs.
To maximize event rates, microlensing searches focus
on monitoring of very crowded, and scientifically attractive,
stellar fields; the Galactic Bulge region and Magellanic Clouds.
Some observations are conducted in denser portions of the Galactic disk.

It is a common situation nowadays that the ability to generate data
far exceeds the ability to process it, and even more so, to comprehend it.
The list of projects which aim at monitoring significant parts of the sky for
variability includes more than 30 names
({\it http://www.astro.princeton.edu/faculty/bp.html}), yet only a small
fraction of those can process the data efficiently enough
to make the measurements publicly available soon after the data is taken
(e.g. Brunner et al. 2001). The issue of exporting the data in a convenient
form compounds the problem. The National Virtual Observatory (NVO)
project has very ambitious plans to provide the tools and some
standards (perhaps de facto standards) for processing the large
amounts of information and web data publication (Szalay 2001).
Large catalogs have added complexity
(project description {\it http://www.us-vo.org/}).
By the time some sort of processing is complete, new information emerges
in the process, frequently information which should be incorporated into
the catalog. It seems that the only static layer is the raw data itself,
typically CCD images, however the photometric output from number crunchers
should also be reasonably slow to change with the new developments.

A regular practice in OGLE is to release the data in the public domain
as soon as possible. The most significant contributions are:
BVI maps of dense stellar regions (Udalski et al. 1998b, 2000a),
Cepheids in Magellanic Clouds (Udalski et al. 1999a, 1999b),
eclipsing variables in the SMC (Udalski et al. 1998a),
catalogs of microlensing events (Udalski et al. 2000b, Wo\'zniak et
al. 2001). Examples from other microlensing teams include samples
of MACHO microlensing events (Alcock et al. 1997a, 1997c)
and selected variable star work from MACHO (1997b) and EROS (Afonso et
al. 1999). In real time detection of microlensing events the main
benefit comes from the follow-up work (e.g. Sackett 2000), in practice
only possible with immediate publication on the WWW. Therefore, all major
microlensing teams (OGLE, EROS, MACHO and MOA) have, or had,
active alert systems.

%
%
%
%
%

\noindent A recent contribution to the publicly available data on
variable stars
is a WWW interface to the MACHO database (Allsman \& Axelrod 2001),
which started with somewhat limited features, but has plans for expansion.
Similar ideas of making evolving catalogs have been discussed within OGLE
for some time now and are motivated by the challenges of data
processing/accessibility. The main objective here is not to make a
potential broad user wait for a long time until the team makes the final
refined product. There is a lot of potential use from the data at all
levels of processing, as demonstrated by the serendipitous recovery of
high proper motion stars (Eyer \& Wo\'zniak 2001) and discovery of the
longest microlensing event ever observed, most likely caused by a black hole
with the mass of several solar masses (Mao et al. 2002), both found in
preliminary OGLE catalogs. OGLE has just released an online catalog
of $\sim$70,000 candidate variables in the LMC and SMC
(\.Zebrun, Soszy\'nski et al. 2001).
With this paper we release an initial catalog of 221,801 candidate
variables in the Galactic Bulge from Difference Image Analysis of
OGLE-II data from seasons 1997-1999. Parts of the current edition are
still not fully calibrated and should not be used in distance
estimates (Section~\ref{sec:phot}).

We restate the basic information about the data in Section~\ref{sec:data}
and in Section~\ref{sec:phot} we briefly summarize the process of finding
variability. Section~\ref{sec:ftp} gives the details of how the
catalog is structured, followed by final remarks and future plans in
Section~\ref{sec:discussion}.

\section{Data}

\label{sec:data}

All OGLE-II frames were collected with the 1.3 m Warsaw Telescope
at the Las Campanas Observatory, Chile. The observatory is operated by the
Carnegie Institution of Washington. The ``first generation'' OGLE camera uses
a SITe $2048 \times 2049$ CCD detector with $24 \mu$m pixels resulting
in 0.417$\arcs$ pixel$^{-1}$ scale. Images of the Galactic bulge are taken in
drift-scan mode at ``medium'' readout speed with a gain of 7.1 $e^-$/ADU
and readout noise of 6.3 $e^-$. The saturation level is about 55,000 ADU.
For details of the instrumental setup, we refer the reader to
Udalski, Kubiak \& Szyma\'nski (1997).

\begin{figure}[Ht]
\vspace{14cm}
\includegraphics{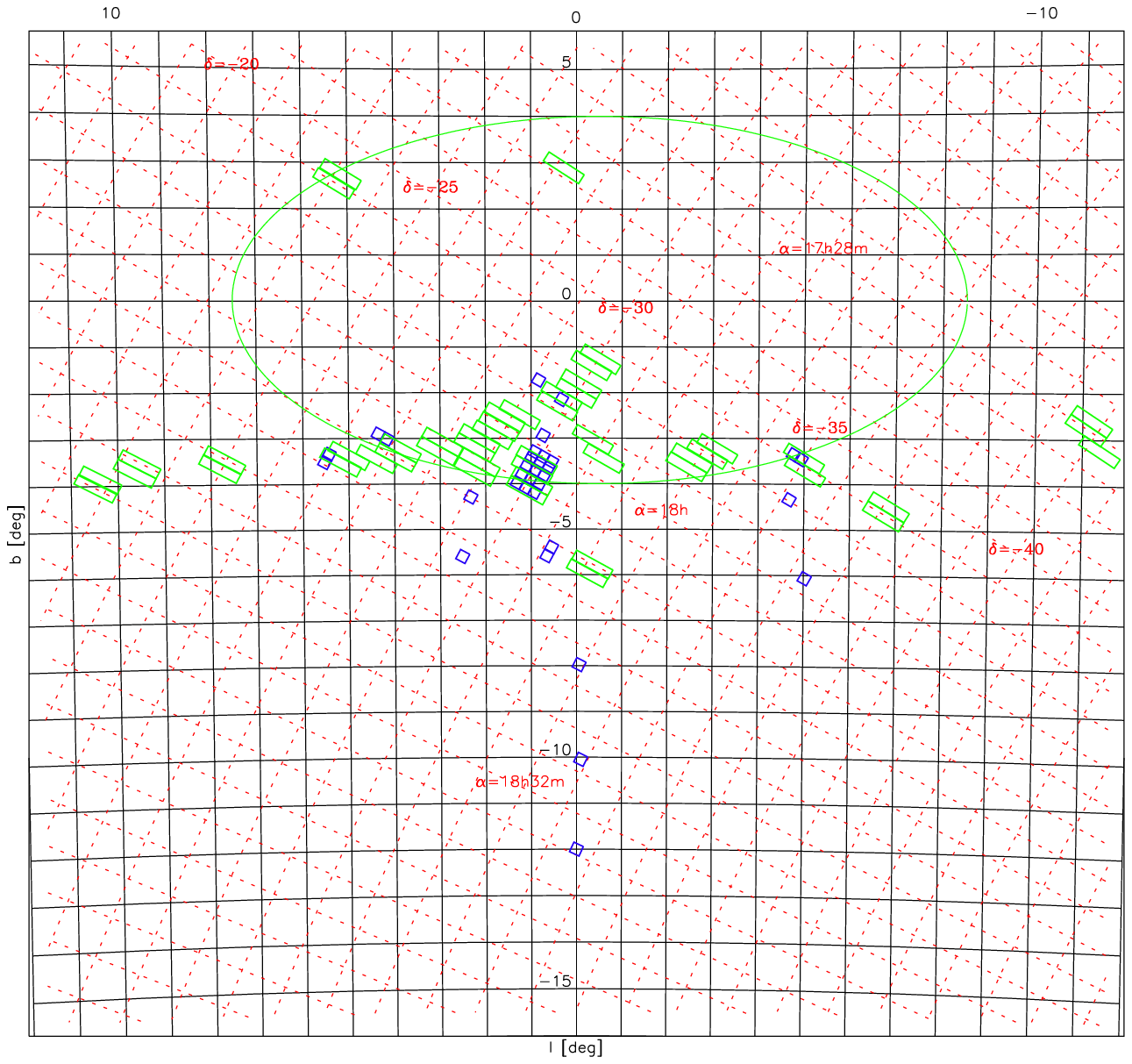}
\caption{\label{fig:fields}
OGLE bulge fields in galactic coordinates (gnomonic projection, great
circles are mapped to straight lines).
Green strips are the OGLE-II scans and blue squares are the old OGLE-I fields.
Large oval indicates the location of the Galactic bar. Fields are selected
in windows of low extinction and avoid very bright foreground stars.
}
\end{figure}

The majority of frames were taken in the $I$ photometric band. The effective
exposure time is 87 seconds. During observing
seasons of 1997--1999 the OGLE experiment typically collected between
200 and 300 $I$-band frames for each of the 49 bulge fields BUL\_SC1--49.
OGLE-II images are 2k$\times$8k strips, corresponding to
$14\arcm\times 57\arcm$ in the sky, therefore the total area of the
bulge covered is about 11 square degrees. The number of frames in $V$ and $B$
bands is small and we do not analyze them with the DIA method. The median
seeing is 1.3$\arcs$ for our data set. In Table~\ref{tab:fields} we provide
equatorial and galactic coordinates of the field centers, the total number
of analyzed frames and the number of candidate variables detected.
Figure~\ref{fig:fields} schematically shows
locations of the OGLE-II bulge fields with respect to the Galactic
bar. Fields BUL\_SC45 and BUL\_SC46 were observed much less
frequently, mostly with the purpose of maintaining phases of variable
stars discovered by OGLE-I. Observations of fields BUL\_SC47--49
started in 1998; the season of 1997 is not available for them.

\section{Extracting variability from OGLE-II bulge frames using DIA}

\label{sec:phot}

The DIA data pipeline we used is based on the image subtraction algorithm
developed by Alard \& Lupton (1998) and Alard (2000), and was written by
Wo\'zniak (2000). Processing of a large 2k$\times$8k pixel frame is performed
after dividing it into 512$\times$128 pixel subframes, with 14 pixel margin
to ensure smooth transitions between coordinate transformations and fits
to spatially variable PSF for individual pieces. Small subframe size allows
us to use polynomial fits for drift-scan images, in which PSF shape and local
coordinate system vary on scales of 100--200 pixels. The reference image,
subtracted from all images of any given field, is a stack of 20 best images
in the sequence.

We adopted kernel expansion used by Wo\'zniak (2000), generally applicable
to all OGLE-II data. The kernel model, represented by a 15$\times$15
pixel raster, consists of 3 Gaussians with sigmas 0.78, 1.35, and 2.34 pixels,
multiplied by polynomials of orders 4, 3, and 2 respectively. The pipeline
delivers a list of candidate variable objects and their difference light
curves. The initial filtering is very weak, with only a minimum of assumptions
made about the variability type. Candidate variables are flagged as
``transient''
or ``continuous'' variables depending on whether variability is confined to
episodes in an otherwise quiet object, or spread throughout the observed time
interval. The total number of candidate variables in all 49 fields was slightly
over 220,000, including 150,000 ``continuous'' and 66,000 ``transient''
cases. Only 4600 objects passed both filters, confirming sensible definitions
of classes. The number of detected variable objects in a given field depends
on the number density of stars, extinction, and number of available
measurements. This ranged from about 800 to over 9000 per
field. The photometry files distributed with this publication do not
contain some of the auxiliary information provided by the
pipeline. In binary format they amount to slightly over 1.3 GB.
Reference images take additional 1.6 GB of storage. We compress the
data when possible.

The error distribution in measurements from our DIA pipeline is nearly
Gaussian with the average scatter only 17\% above the Poisson limit for faint
stars near $I$=17--19 mag, gradually increasing for brighter stars,
and reaching 2.5 times photon noise at $I\sim11$ mag (about 0.5\% of the total
flux). Error bars adopted in this paper are photon noise estimates
renormalized using the curve of Wo\'zniak (2000). As the method
effectively monitors all pixels, variable objects may be discovered
even where no object is detected in the reference image. In regular searches
monitoring is conducted only for objects detected in a single good
quality image, a template. This issue is related to centroid finding.
Currently in our DIA pipeline centroids are calculated based on the
variable signal in a number of frames. As a result the centroid will
be poorly known for an object with low S/N variability, even if it is
very bright on the reference image. Ideally one would use both pieces
of information. It is usually obvious how to determine the centroid
in the presence of blending when confronted with one particular object
of interest, but an optimal algorithm for extracting all variability
in the field using DIA is yet to be developed.

Difference fluxes were converted to magnitudes using reference flux
values obtained from DoPhot photometry on reference frames. The
process of matching units was identical to that in Wo\'zniak (2000).
DIA observations were tied to the OGLE database of regular PSF
photometry. Most fields were calibrated to 0.05 mag accuracy,
however at the time of this analysis for 10 fields
(BUL\_SC: 7, 9, 20, 25, 28, 32, 43, 47, 48 and 49) only rough calibration
was available and the zero point differences may reach $\pm$0.25 mag.
The catalog will be re-calibrated after merging with the data for
the 2000 observing season.

The conversion is given by the formula

$$ m_I = m_{\rm 0} - 2.5\log(f + f_{\rm ref}), $$

\noindent where $f$ is the difference flux of a particular observation,
$f_{\rm ref}$ is the reference flux, $m_{\rm 0}$ is the magnitude zero point
for a given subframe of the reference image, and $m_I$ is the
converted $I$-band magnitude. All quantities in the formula are
available in the catalog and light curve files (Section~\ref{sec:ftp}).
Due to noise, occasionally one runs into a problem of negative fluxes
in DIA. For those measurements the difference flux may still be
perfectly valid, the magnitude will have an error code and the
appropriate flag will be set (Section~\ref{sec:ftp} and Appendices).

Julian dates of individual observations also bear some discussion.
In drift-scan observing the time of mid exposure depends on the
position of the object. In the case of the Galactic Bulge data of OGLE-II,
the correction which should be added to the starting time of the scan
is given by:

$$dt = ( Y + 1024 )\times 0.0423816/86400~~~~~~HJD   < 2451040~$$
$$dt = ( Y + 1024 )\times 0.0484627/86400~~~~~~HJD \ge 2451040,$$

\noindent
where $Y$ is the pixel position in the reference image along the direction
of the scan, in the range 0--8192 in OGLE-II, resulting in differences
reaching several minutes. The time stamps of observations in the
catalog have been corrected for this effect (Section~\ref{sec:ftp}).

\section{FTP catalog}

\label{sec:ftp}

The catalog of candidate bulge variables presented in this paper
is available via FTP from
{\it ftp://bulge.princeton.edu/ogle/ogle2/bulge\_dia\_variables}.
The data is naturally divided into 49 parts for fields BUL\_SC1 -- BUL\_SC49.
Reference images for each field (2k $\times$ 8k FITS frames) are stored
in the subdirectory {\tt reference\_frames}. Information is available in
two formats: plain text (subdirectory {\tt plain\_text}) and binary FITS tables
(subdirectory {\tt fits\_tables}). FITS format is an astronomical standard
and the ease of its use with programs like IDL is remarkable. There
are two types of files for each field: the catalog of candidate
variables, and the database of light curves. The catalog contains a single
entry per object with the overall parameters of the light curve and
identifying information, like coordinates. Below is a sample record
with the explanation of fields:

3764~~~~~207.14~~6721.60~~~~~271.272315~~$-$28.578460~~~~~18:05:05.35~~$-$28:34:42.5

17.407~~0.783~~~~~359.5~~23.38~~0.13~~~~~1~~152~~0~~181~~181~~1

\begin{enumerate}

\item $\tt [~~~~~]$ --- number of candidate variable as returned
                        by the pipeline

\item $\tt [X\_TPL]$ --- x template coordinate 
	(0.0 is the middle of the bottom left pixel)

\item $\tt [Y\_TPL]$ --- y template coordinate 
	(0.0 is the middle of the bottom left pixel)

\item $\tt [RA]$ --- RA in decimal degrees

\item $\tt [DEC]$ --- DEC in decimal degrees

\item $\tt [RA\_STR]$ --- RA in sexagesimal hours

\item $\tt [DEC\_STR]$ --- DEC in sexagesimal degrees

\item $\tt [MEAN\_MAG]$ --- mean of all magnitude values
				which could be determined

\item $\tt [MAG\_SCAT]$ --- scatter of all magnitudes used in mean calculation

\item $\tt [REF\_FLUX]$ --- reference flux

\item $\tt [MAG\_0]$ --- magnitude zero point

\item $\tt [ID\_RAD]$ --- distance between the centroid of the variable
	and the nearest DoPhot star in the reference image in pixels
        (for the calculation of the reference flux and conversion
        to magnitudes)

\item $\tt [VTYPE]$ --- type of variable coded in bits of a 2 byte integer:
	1st bit -- ``transient'', 2nd bit -- ``continuous''. Therefore
        the value of the integer will be 1 for ``transient'',
        2 for ``continuous variable '', and 3 for both
	(see Section~\ref{sec:phot} for details).

\item $\tt [N\_FRAMES]$ --- number of frames used in centroid determination

\item $\tt [N\_BAD]$ --- number of bad pixels in the fitting radius
			on the reference image

\item $\tt [NGOOD]$ --- number of ``good'' flux measurements. A ``good'' point
      is the one for which none of the several
      types of problems monitored by the pipeline occurred
      (flags 1--10 in Appendix~B are set to 0).

\item $\tt [NMAG]$ --- number of magnitude values which could be determined
      ( the ones which are not determined come from non-positive fluxes)

\item $\tt [FLAG]$ --- flags, see the explanation below

\end{enumerate}

Several kinds of problematic situations are reported as flags in the
last column of the catalog file. Flags are explained in Appendix~A.

Capitalized names after the column number are names of columns in
binary FITS tables ({\tt bul\_sc*\_cat.fts}). An empty bracket means
that this column is omitted in the FITS table, but it is present in
the text file. In text version of these
files ({\tt bul\_sc*\_cat.dat}) columns have no names and are identified
by their order. The database of light curves includes all measurements
for all detected objects. Light curves in plain text format are stored
one per file and grouped by the field. For example, subdirectory
{\tt BUL\_SC1} in {\tt plain\_text} contains 4597 {\tt bul\_sc1\_*.dat.gz} 
files, compressed to save space and transfer time. The columns in light curve
files are as follows:

\begin{enumerate}

\item $\tt [OBS\_TIME]$ --- Heliocentric Julian Day of the observation,
offset by 2450000.0

\item $\tt [DIFF\_FLUX]$ --- difference flux ($-$99.00 for error code)

\item $\tt [FLUX\_ERR]$ --- difference flux error ($-$99.00 for error code)

\item $\tt [MAG]$ --- $I$ band magnitude ($-$1.0 for error code)

\item $\tt [MAG\_ERR]$ --- $I$ band magnitude error ($-$1.0 for error code)

\item $\tt [FLAG]$ --- flags, explained below

\end{enumerate}

The flags are explained in Appendix~B. They provide a lot of information
on whether the measurement is valid or not and common problems which
may have affected its reliability. To stay on the conservative side,
only measurements with no flags should be used (integer value 0).
In {\tt plain\_text} subdirectory there are also 49 {\tt bul\_sc*\_db.tar}
files with all light curve files for a given field grouped together
for convenient transfers. 

In binary FITS format all light curves for each field are stored in a
single table. The first extension contains the names of the frames and
starting times of the drift-scan exposures in Heliocentric Julian Days
shifted by 2450000.0. These time stamps are identical for all objects
in a single image. However, the effective time of mid exposure varies
depending on the position of the object along the scan. The corrected
times of observations are provided for each star separately in the second
FITS extension with the actual photometry (Section~\ref{sec:phot}).
All measurements for all stars are stored in the same columns and
identified by their index within the column. The number of
observations per star is fixed and given by the length of the time
vector from the first extension. Ordering is such that the number of
the individual observation within a single light curve is ascending
fastest along the column of the binary table. If, for example, the
number of dates in the first extension of {\tt bul\_sc1\_db.fts} is
197, the first 197 rows of the second FITS extension correspond to the
first light curve, the next 197 rows are the second light curve and so
on. The total number of rows is 197 $\times$ 4597, where 4597
is the number of candidate variables in BUL\_SC1 field, the same as the number
of rows in catalog files {\tt bul\_sc1\_cat.dat} and {\tt bul\_sc1\_cat.fts}.
This information, along with several other useful numbers, is stored
in headers.

In Appendix~C we include the explanation and values of the pipeline
parameters which were important for detection of variables.

\section{Discussion and future work}

\label{sec:discussion}

As mentioned before, the current edition of the catalog basically includes
entire output of the DIA pipeline as described by Wo\'zniak (2000),
supplemented with determinations of the reference flux to put the
light curves on the magnitude scale. Some optimization has been performed
to keep the contamination by artifacts low without rejecting too many
real variable objects, but it must be clearly stated that about 10\%
of the light curves in the present release are not real objects and
result from various problems, undetected at the pipeline level.
We are extending the work of Mizerski \& Bejger (2001) from the first
BUL\_SC1 field to all fields in the effort to flag several common
types of artifacts and clean the sample of spurious objects. 

Classifying a real variable star versus a spurious one is the first step
in the interpretation of this data. Ultimately we envision increasingly
refined information added to the catalog to facilitate applications.
This should include the full classification of the detected variables,
cross identification with objects found by 2MASS and in X-ray
catalogs, periods for periodic sources etc. The work on automated
classification of periodic variables is in progress. In addition to
examining 2-D projections of a multidimensional parameter space and
trying to code a human made algorithm (see Mizerski \& Bejger
2001 and Wo\'zniak et al. 2001 for such work on this data), we are
experimenting with data mining techniques. Even with the current volume
of data in OGLE-II we believe it is enabling to make the transition
from ''telling the computer how to do it'' to ''telling the computer
what to do'' and leaving the rest to the algorithm. A number of
standard machine learning tools are available, which take small
preclassified subsets of light curves and can ''learn'' to classify
the rest of the data.

\.Zebrun, Soszy\'nski et al. (2001) provided a convenient web interface
to access the data on variables in the LMC and SMC. It is our
intention to build a similar tool with the addition of positional
searches. Although the volume of the data which can be accessed by
browsing web pages is limited in practice, the search by coordinates
is a powerful tool for numerous applications. Transfer by FTP and
distribution of DAT tapes are currently primary modes of accessing
major parts of this catalog. For your copy of a DAT tape, please
contact Prof. Bohdan Paczy\'nski (email: bp@astro.princeton.edu,
mail: Princeton University Observatory, Princeton, NJ, 08544).
To access this archive online use the OGLE web site
{\it http://bulge.princeton.edu/$\sim$ogle/ogle2/bulge\_dia\_variables} .

\acknowledgements

We thank Prof. Paczy\'nski for support and encouragement in this project.
This work was supported by the NSF grant AST-9820314 to B. Paczy\'nski,
and Polish KBN grant 2P03D01418 to M. Kubiak.
Additional support for P. Wo\'zniak was provided under
the DOE contract W-7405-ENG-36.


\clearpage

\begin{deluxetable}{lccrrrr}
\tablecaption{\label{tab:fields}{OGLE-II bulge fields.}}
\tablewidth{12.5cm}
\tablehead{
\colhead{Field} &
\colhead{$\alpha_{2000}$} &
\colhead{$\delta_{2000}$} &
\colhead{$l$} &
\colhead{$b$} &
\colhead{$N_{\rm obs}$} &
\colhead{$N_{\rm var}$}
\nl
\colhead{} &
\colhead{$\uph~~\upm~~\ups$} &
\colhead{$\arcd~~\arcm~~\arcs$} &
\colhead{$\arcd$} &
\colhead{$\arcd$} &
\colhead{} &
\colhead{}
}
\startdata
 BUL\_SC1 & 18:02:32.5 & $-$29:57:41 &    1.08 &  $-$3.62 &  197 & 4597 \\
 BUL\_SC2 & 18:04:28.6 & $-$28:52:35 &    2.23 &  $-$3.46 &  192 & 5279 \\
 BUL\_SC3 & 17:53:34.4 & $-$29:57:56 &    0.11 &  $-$1.93 &  309 & 8493 \\
 BUL\_SC4 & 17:54:35.7 & $-$29:43:41 &    0.43 &  $-$2.01 &  324 & 9096 \\
 BUL\_SC5 & 17:50:21.7 & $-$29:56:49 & $-$0.23 &  $-$1.33 &  307 & 7257 \\
 BUL\_SC6 & 18:08:03.7 & $-$32:07:48 & $-$0.25 &  $-$5.70 &  228 & 3211 \\
 BUL\_SC7 & 18:09:05.5 & $-$32:07:10 & $-$0.14 &  $-$5.91 &  219 & 1618 \\
 BUL\_SC8 & 18:23:06.2 & $-$21:47:53 &   10.48 &  $-$3.78 &  211 & 2331 \\
 BUL\_SC9 & 18:24:00.0 & $-$21:47:10 &   10.59 &  $-$3.98 &  212 & 1847 \\
BUL\_SC10 & 18:20:06.6 & $-$22:23:03 &    9.64 &  $-$3.44 &  220 & 2499 \\
BUL\_SC11 & 18:21:06.5 & $-$22:23:05 &    9.74 &  $-$3.64 &  215 & 2256 \\
BUL\_SC12 & 18:16:06.3 & $-$23:57:54 &    7.80 &  $-$3.37 &  209 & 3476 \\
BUL\_SC13 & 18:17:02.6 & $-$23:57:44 &    7.91 &  $-$3.58 &  208 & 3084 \\
BUL\_SC14 & 17:47:02.7 & $-$23:07:30 &    5.23 &     2.81 &  209 & 4051 \\
BUL\_SC15 & 17:48:06.9 & $-$23:06:09 &    5.38 &     2.63 &  204 & 3853 \\
BUL\_SC16 & 18:10:06.7 & $-$26:18:05 &    5.10 &  $-$3.29 &  202 & 4802 \\
BUL\_SC17 & 18:11:03.6 & $-$26:12:35 &    5.28 &  $-$3.45 &  200 & 4690 \\
BUL\_SC18 & 18:07:03.5 & $-$27:12:48 &    3.97 &  $-$3.14 &  203 & 5805 \\
BUL\_SC19 & 18:08:02.4 & $-$27:12:45 &    4.08 &  $-$3.35 &  195 & 5255 \\
BUL\_SC20 & 17:59:16.0 & $-$28:52:10 &    1.68 &  $-$2.47 &  235 & 5910 \\
BUL\_SC21 & 18:00:22.3 & $-$28:51:45 &    1.80 &  $-$2.66 &  238 & 7449 \\
BUL\_SC22 & 17:56:47.6 & $-$30:47:46 & $-$0.26 &  $-$2.95 &  275 & 5589 \\
BUL\_SC23 & 17:57:54.5 & $-$31:12:36 & $-$0.50 &  $-$3.36 &  255 & 4815 \\
BUL\_SC24 & 17:53:17.9 & $-$32:52:45 & $-$2.44 &  $-$3.36 &  250 & 4304 \\
BUL\_SC25 & 17:54:21.0 & $-$32:52:10 & $-$2.32 &  $-$3.56 &  243 & 3046 \\
BUL\_SC26 & 17:47:15.5 & $-$34:59:31 & $-$4.90 &  $-$3.37 &  241 & 4713 \\
BUL\_SC27 & 17:48:23.6 & $-$35:09:32 & $-$4.92 &  $-$3.65 &  220 & 3691 \\
BUL\_SC28 & 17:47:00.0 & $-$37:07:10 & $-$6.76 &  $-$4.42 &  217 & 1472 \\
BUL\_SC29 & 17:48:10.8 & $-$37:07:21 & $-$6.64 &  $-$4.62 &  211 & 2398 \\
BUL\_SC30 & 18:01:25.0 & $-$28:49:55 &    1.94 &  $-$2.84 &  233 & 6893 \\
BUL\_SC31 & 18:02:22.6 & $-$28:37:21 &    2.23 &  $-$2.94 &  236 & 4789 \\
BUL\_SC32 & 18:03:24.0 & $-$28:37:10 &    2.34 &  $-$3.14 &  231 & 5007 \\
BUL\_SC33 & 18:05:30.9 & $-$28:52:50 &    2.35 &  $-$3.66 &  187 & 4590 \\
BUL\_SC34 & 17:58:18.5 & $-$29:07:50 &    1.35 &  $-$2.40 &  239 & 7953 \\
BUL\_SC35 & 18:04:28.6 & $-$27:56:56 &    3.05 &  $-$3.00 &  177 & 5169 \\
BUL\_SC36 & 18:05:31.2 & $-$27:56:44 &    3.16 &  $-$3.20 &  209 & 8805 \\
BUL\_SC37 & 17:52:32.2 & $-$29:57:44 &    0.00 &  $-$1.74 &  305 & 8367 \\
BUL\_SC38 & 18:01:28.0 & $-$29:57:01 &    0.97 &  $-$3.42 &  191 & 5072 \\
BUL\_SC39 & 17:55:39.1 & $-$29:44:52 &    0.53 &  $-$2.21 &  318 & 7338 \\
BUL\_SC40 & 17:51:06.1 & $-$33:15:11 & $-$2.99 &  $-$3.14 &  205 & 4079 \\
BUL\_SC41 & 17:52:07.2 & $-$33:07:41 & $-$2.78 &  $-$3.27 &  208 & 4035 \\
BUL\_SC42 & 18:09:05.0 & $-$26:51:53 &    4.48 &  $-$3.38 &  204 & 4360 \\
BUL\_SC43 & 17:35:10.0 & $-$27:10:10 &    0.37 &     2.95 &  251 & 3351 \\
BUL\_SC44 & 17:49:22.4 & $-$30:02:45 & $-$0.43 &  $-$1.19 &  258 & 7836 \\
BUL\_SC45 & 18:03:33.0 & $-$30:05:00 &    0.98 &  $-$3.94 &   88 & 2262 \\
BUL\_SC46 & 18:04:36.0 & $-$30:05:00 &    1.09 &  $-$4.14 &   84 & 2057 \\
BUL\_SC47 & 17:27:00.0 & $-$39:46:00 &$-$11.19 &  $-$2.60 &  151 & 1152 \\
BUL\_SC48 & 17:28:10.0 & $-$39:46:00 &$-$11.07 &  $-$2.78 &  145 &  973 \\
BUL\_SC49 & 17:29:20.0 & $-$40:16:00 &$-$11.36 &  $-$3.25 &  142 &  826 \\

\enddata
\end{deluxetable}

\clearpage

\clearpage

\section{APPENDIX A}

Catalog flags are coded as single bits of a 4 byte integer and
listed below (the least significant bit first). Integer value 12, e.g.,
means that flags 3 and 4 are true and the rest are false. The values
quoted for selected pipeline parameters have been actually used
to set the flags in this analysis.

\begin{enumerate}

\item crowding flag, set if within $\pm4$ pixels of the maximum
      pixel with flux $f_0$ there is a secondary local maximum with
      pixel flux $f > 0.15\times f_0\times r$, where $r$ is the
      distance from the star centroid in pixels

\item fewer than ${\tt N\_FRAMES = 4}$ used in centroid finding

\item more than ${\tt N\_BAD = 0}$ bad pixels on the reference image
      within the fitting radius of 3.0 pixels

\item fraction of less than ${\tt MIN\_GFRA = 0.5}$ difference flux
      measurements are ``good'' from the total number of frames taken
      for the field. A ``good'' point is the one for which none of the several
      types of problems monitored by the pipeline occurred
      (flags 1--10 in Appendix~B are set to 0).

\item mean magnitude and its scatter could not be calculated
	because fewer than 2 individual magnitudes were defined

\end{enumerate}

\clearpage

\section{APPENDIX B}

Light curve flags are coded as single bits of a 4 byte integer and
listed below (the least significant bit first). Integer value 12, e.g.,
means that flags 3 and 4 are true and the rest are false. The values
quoted for selected pipeline parameters have been actually used
to set the flags in this analysis.

\begin{enumerate}

\item pipeline returned error code for difference flux

\item pipeline returned error code for flux error

\item ${\chi^2}$ per pixel of the difference subframe larger than
      ${\tt MAXCHI2I = 6.0}$

\item ${\chi^2}$ per pixel of the PSF fit larger than ${\tt MAXCHI2N = 1.0e32}$
      (effectively no cut)
      
\item FWHM of the PSF fit larger than ${\tt MAX\_FWHM = 3.4}$ pix

\item number of bad pixels within the fitting radius larger than
      ${\tt MAX\_NBAD = 3}$

\item correlation coefficient with the PSF lower than
      ${\tt MIN\_CORR = 0.0}$ (effectively no cut)

\item star in the rejected region of the CCD (currently empty)
	
\item flux error ${\tt NSIGERR = 10}$ times larger than percentile
      ${\tt ERRFRAC = 0.5}$ of all individual flux errors
      (0.5 corresponds to median)

\item ${\chi^2}$ per pixel of the PSF fit ${\tt NSIGCHI2 = 10}$ times
      larger than
      percentile ${\tt CHI2FRAC = 0.5}$ of all individual values
      (0.5 corresponds to median)

\item magnitude could not be calculated due to missing or non-positive fluxes

\item magnitude error could not be calculated due to missing or non-positive
      values

\end{enumerate}

\clearpage

\section{APPENDIX C}

Explanation of light curve cleaning parameters. The values quoted
have been used to set the flags in Appendices A and B.

\begin{enumerate}

\item ${\tt [MAXNMAD0 = 0]}$ --- max number of bad pixels on the reference
      image within the fitting radius

\item ${\tt [MINNFRM0 = 4]}$ --- min number of frames used in centroid
      calculation 

\item ${\tt [MAX\_NBAD = 3]}$ --- max number of bad pixels on a given image
      within the fitting radius

\item ${\tt [MIN\_GFRA = 0.5]}$ --- min fraction of good points within entire
      sequence of frames 

\item ${\tt [BAD\_FLUX = -99.0]}$ --- error code for difference flux 

\item ${\tt [BAD\_ERR = -99.0]}$ --- error code for flux error

\item ${\tt [MAXCHI2N = 1.0e32]}$ --- max ${\chi^2}$ per pixel for PSF fit 

\item ${\tt [MAXCHI2I = 6.0]}$ --- max ${\chi^2}$ per pixel for difference
                                 subframe

\item ${\tt [MIN\_CORR = 0.0]}$ --- min correlation coefficient with the PSF

\item ${\tt [MAX\_FWHM = 3.4]}$ --- max FWHM in pixels

\item ${\tt [ERRCLN = 1]}$ --- is flagging of large error bars on ? (1 = yes)

\item ${\tt [NSIGERR = 10.0]}$ --- base threshold is multiplied by
      this factor to get the final threshold for error bar

\item ${\tt [ERRFRAC = 0.5]}$ --- percentile of the error distribution
      for base threshold

\item ${\tt [CHI2CLN = 1]}$ --- is flagging poor PSF fits on ? (1 = yes )

\item ${\tt [NSIGCHI2 = 10.0]}$ --- base threshold is multiplied by
      this factor to get the final threshold for ${\chi^2}$ of the PSF fit

\item ${\tt [CHI2FRAC = 0.5]}$ --- percentile of the ${\chi^2}$ per pix
      distribution for base threshold

\end{enumerate}

\end{document}